\documentclass[cameraready]{Interspeech}

\usepackage{amsmath,graphicx}
\usepackage{amsfonts}
\usepackage{booktabs}
\usepackage[subrefformat=parens]{subcaption}
\usepackage{siunitx}
\usepackage{mathtools}
\usepackage{comment}
\usepackage{multirow}
\usepackage{makecell}
\usepackage{threeparttable}
\usepackage{arydshln}   
\usepackage[normalem]{ulem} 
\usepackage{cite}
\usepackage{amsthm}
\usepackage[ruled,linesnumbered,vlined]{algorithm2e}
\usepackage{bbm}    
\usepackage{inconsolata}
\usepackage{mleftright}
\mleftright

\def\equationautorefname~#1\null{(#1\null)}

\renewcommand{\sectionautorefname}{Section}
\renewcommand{\subsectionautorefname}{\sectionautorefname}

\let\orgautoref\autoref
\providecommand{\Autoref}[1]
{%
\def\figureautorefname{Figure}%
\def\subfigureautorefname{Figure}%
\orgautoref{#1}%
}
\renewcommand{\autoref}[1]
{%
\def\figureautorefname{Fig.}%
\def\subfigureautorefname{\figureautorefname}%
\def\sectionautorefname{Sec.}%
\def\subsectionautorefname{\sectionautorefname}%
\def\subsectionautorefname{\sectionautorefname}%
\orgautoref{#1}%
}

\newcommand{\vect}[1]{\mbox{\boldmath $#1$}}
\newcommand{\abs}[1]{\left\lvert#1\right\rvert}
\newcommand{\norm}[1]{\left\lVert#1\right\rVert}

\newcommand{\trans}[1]{#1^\mathsf{T}}

\DeclareMathOperator*{\argmin}{arg\,min}\def\appendixautorefname~#1\null{~#1 \null}

\DeclareFontEncoding{LS1}{}{}
\DeclareFontSubstitution{LS1}{stix}{m}{n}
\DeclareSymbolFont{stixletters}{LS1}{stix}{m}{it}
\DeclareMathAccent{\anticausal}{\mathord}{stixletters}{"91}
\DeclareMathAccent{\causal}{\mathord}{stixletters}{"92}
\DeclareMathAccent{\bidirectional}{\mathord}{stixletters}{"95}
\DeclareMathAccent{\noncausal}{\mathord}{stixletters}{"95}



\makeatletter
   {\list{}{\leftmargin=10pt 
            \labelwidth\z@ \itemindent-\leftmargin
            }}%
   {\endlist}
\makeatother

\makeatletter
\patchcmd{\@algocf@start}
  {-1.5em}
  {0pt}
{}{}
\makeatother
\appto\TPTnoteSettings{\footnotesize}

\title{\texorpdfstring{Tight Boundary Prediction in Speaker Diarization Using\\Causal--Anticausal Consistency}{Tight Boundary Prediction in Speaker Diarization Using Causal--Anticausal Consistency}}

\author[orcid=0000-0002-3166-4956, correspondingauthor]{Shota}{Horiguchi}
\author[orcid=0000-0002-5175-7834]{Marc}{Delcroix}
\author[orcid=0000-0002-1130-5059]{Naohiro}{Tawara}
\author[orcid=0009-0003-4322-4127]{Takanori}{Ashihara}
\author[orcid=0000-0002-3971-0654]{Atsushi}{Ando}

\address{
    NTT, Inc., Japan
}

\email{horiguchi@ieee.org}


\begin{document}
\abovedisplayskip=2pt
\belowdisplayskip=2pt

\setlength\textfloatsep{7pt}
\setlength\dbltextfloatsep{7pt}
\setlength\floatsep{7pt}
\setlength\abovecaptionskip{2pt}
\setlength\belowcaptionskip{2pt}
\captionsetup[subfloat]{aboveskip=2pt,belowskip=8pt}

\maketitle

\begin{abstract}
Multi-talker conversational automatic speech recognition data are often used to train speaker diarization models.
Because such data prioritize semantic continuity, pauses and boundary margins are included within speech segments, resulting in loose annotations.
Models trained on such data tend to internalize mechanisms that reproduce this looseness, although tight speech intervals are sometimes preferable for downstream applications.
In this paper, we address the novel task of enabling models to produce tight predictions using loose labels.
Our method generates tighter pseudo labels using causal and anticausal models, which are inherently incapable of learning loosening behavior.
We further propose a co-training scheme that iteratively tightens labels and updates both models for more progressive refinement.
Experimental results show that the proposed method recovers about \SI{70}{\percent} of the tightening effect achieved by ideal tight-label training and improves downstream performance.
\end{abstract}

\section{Introduction}
Speaker diarization is the task of determining who is speaking when from an audio signal.
It is typically addressed using models that take audio as input and output speaker-wise speech segments~\cite{fujita2019end1,fujita2019end2,medennikov2020targetspeaker,kinoshita2021integrating,horiguchi2022encoderdecoder,wang2023target,harkonen2024eend,cheng2025sequence}.
Training such models requires a large amount of data annotated with speaker-wise speech intervals, and in practice, multi-talker automatic speech recognition (ASR) corpora are often used for this purpose~\cite{bredin2023pyannote,plaquet2023powerset,han2025leveraging,han2025fine}.

\begin{figure}[t]
\includegraphics[width=\linewidth]{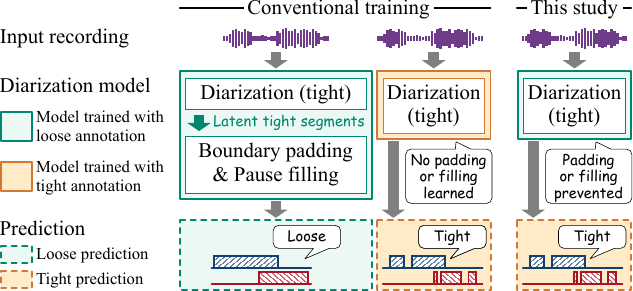}\vspace{0.3em}
\caption{Comparison of model behavior in conventional training and in this study.}\label{fig:task}
\end{figure}

A recent study has shown that speech segments in ASR corpora are often loosely annotated~\cite{horiguchi2025can}: pauses within utterances are labeled as part of speech, and segment boundaries are padded at both ends to prevent truncation~\cite{fu2021aishell,carletta2007unleashing,yu2022m2met,barker2018fifth,van2020dipco}.
When diarization models are trained with such loose labels, their outputs also tend to be loose (\autoref{fig:task}, left).
This property is not always desirable for applications that rely on diarization results.
For example, in guided source separation (GSS)~\cite{boeddeker2018front,raj2023gpu}, where diarization results are used as prior knowledge to estimate time-frequency masks, tighter intervals provide stronger constraints on speech absence.
Speaker diarization or channel-wise voice activity detection (VAD) is also used to generate training data for spoken dialogue models~\cite{defossez2024moshi,fu2024investigating,ohashi2025towards} and to evaluate the naturalness of conversational statistics in generated data~\cite{veluri2024beyond,cui2026turnguide}.
In these cases, loose outputs may blur conversational structure: pauses may be classified as parts of speech intervals, or backchannel may be treated as part of interruptions.
Therefore, there is a need for speaker diarization models that can produce tighter speech intervals.

At present, the only way to train a model that can predict segments with tight boundaries is to provide tight labels as supervision (\autoref{fig:task}, center).
In the prior study~\cite{horiguchi2025can}, forced alignment was used to tighten loose segments in ASR corpora; however, this approach requires recordings in which each speaker is recorded on a separate channel, such as headset recordings.
For audio uploaded to sharing platforms, which are fundamental data sources for training high-performing models nowadays~\cite{radford2023robust,ghosh2025audio}, all speakers’ speech is mixed into a single channel, making this method inapplicable.
On the other hand, manual tight annotation is costly.
One dataset series that provides such tight annotations is DIHARD I--III~\cite{ryant2018first,ryant2019second,ryant2021third}.
According to the DIHARD III evaluation plan~\cite{ryant2020third}, \textit{
it was found that
manual annotation to this spec 
[...] required use of highly skilled and experienced annotators 
[...] making the annotation extremely slow and costly 
[...] with real time rates typically greater than 15X and sometimes exceeding 30X}.
Another example is VoxConverse~\cite{chung2020spot}, which provides annotations for YouTube videos.
It accelerates annotation by leveraging face-tracking results from video data.
However, this approach assumes videos in which speakers are clearly visible and still ``\textit{takes around twice the video duration}'' to annotate.
Paying this level of additional cost on top of transcription expenses is not practical. 

In this paper, we address a novel task: training a model to output tight segments from loose annotations in ASR corpora (\autoref{fig:task}, right).
In conventional diarization model training, loose target labels result in similarly loose outputs (\autoref{fig:task}, left).
This occurs because the model becomes entangled not only with detecting underlying tight speech segments but also with padding segment boundaries and filling pauses.

To address this task, we propose a method that generates tighter pseudo labels from the outputs of causal and anticausal models and uses them to train a final non-causal model.
Even when trained with loose labels, causal and anticausal models are inherently incapable of learning to pad segment boundaries or fill pauses.
A causal model cannot pad before a speech segment, whereas an anticausal model cannot pad after it.
Moreover, neither model can immediately determine whether a silence following a speech segment corresponds to an actual termination or a pause to be filled.
As a result, their capability to reproduce loose boundaries is inherently limited compared with that of a non-causal model.
Accordingly, we use their outputs to mask loose annotations and generate tighter pseudo labels.
We further propose the co-training of causal and anticausal models.
At each training step, the labels are tightened using their outputs and immediately used for parameter updates, thereby enabling progressive tightening.
Finally, we empirically demonstrate that a non-causal model trained with the resulting tighter pseudo labels produces tighter predictions than one trained with loose labels, leading to improvements in downstream tasks.

\section{Related work}
\subsection{Speaker diarization}
Training end-to-end neural diarization (EEND) models requires a large-scale dataset.
In early studies, since it was difficult to achieve large-scale training using only real data, simulated mixtures were used for pretraining~\cite{fujita2019end1,kinoshita2021integrating,horiguchi2022encoderdecoder,wang2023target}.
In recent years, an increasing number of corpora have been developed for multi-talker ASR; as a result, it has become possible to aggregate such corpora and combine them with corpora curated for diarization, thereby constructing large-scale datasets composed entirely of real recordings for training~\cite{bredin2023pyannote,plaquet2023powerset,han2025leveraging,han2025fine,horiguchi2025pretraining,han2025efficient}.
At the same time, evaluation has also been conducted on these corpora using the same parameters to demonstrate that the models can generalize across a wide range of domains without adaptation to specific domains~\cite{han2025leveraging,han2025fine,han2025efficient}.

However, a recent study identified issues with training on a compound dataset of ASR and diarization corpora~\cite{horiguchi2025can}.
The study pointed out that speech boundaries in ASR corpora are generally looser than those in diarization corpora.
This is because ASR corpora tend to avoid splitting segments at short pauses, so as not to break sentences in the middle, or truncate speech at segment boundaries.
When an EEND model is trained on a compound dataset with mixed boundary precision, it leads to corpus-specific memorization of boundary tightness.
This behavior is undesirable for real-world applications, since it becomes unclear whether the model will output tight or loose boundaries for data from unseen domains.
It has also been empirically shown that tight boundaries are beneficial in backend applications such as speech separation and ASR, especially when combined with subsequent boundary-loosening operations such as morphological closing.
In light of these facts, it is preferable for speaker diarization models to predict tight speech segments that exclude silent intervals.
The only proposed solution in the aforementioned study was to tighten all labels in the compound dataset via forced alignment, which limits its applicability to corpora with speaker-wise channel-separated recordings.
In contrast, this paper addresses a more challenging problem: enabling EEND models to produce tight boundaries using only loose labels with single-channel mixture recordings.

\subsection{Training from unreliable labels}
The problem we address is that segments annotated as speech may in fact contain silent portions.
This is related to the broader literature on machine learning with noisy labels.
The most common setting assumes that labels are incorrect with some probability due to unreliable annotators.
Typical approaches include the use of noise transition matrices~\cite{natarajan2013learning,patrini2017making} and label smoothing~\cite{lukasik2020does}.
A more realistic setting assumes that the errors are more likely for hard samples.
Representative methods include sample reweighting~\cite{ren2018learning}, curriculum learning~\cite{jiang2018mentornet,guo2018curriculumnet}, and co-training~\cite{han2018co,yu2019does}, which select high-confidence samples based on the probabilities output from the prediction models.
In the speaker diarization and segmentation domain, collar-aware training has been proposed to address annotation noise~\cite{zeghidour2021dive,kalda2022collar}.
However, this approach simplifies the problem by allowing the model to ignore predictions near segment boundaries and is therefore not aligned with our objective, which is to accurately predict those boundaries instead.

What fundamentally distinguishes our problem from these settings is that both ASR-oriented (i.e., loosely annotated) labels and diarization-oriented (i.e., tightly annotated) labels are valid when viewed from their respective perspectives.
It has been experimentally shown that models can learn the characteristics of either labeling scheme and generalize well~\cite{horiguchi2025can}.
From a diarization-oriented perspective, the errors in ASR-oriented labels, which are mostly false alarms, tend to be distributed around true utterance boundaries.
When trained with ASR-oriented labels, a model learns to confidently predict speech in such intervals by leveraging contextual information, even when no speech is actually present, as illustrated in \autoref{fig:task}.
Therefore, the aforementioned confidence-based noisy-label training approaches are not applicable in this setting.

\begin{figure*}[t]
\subfloat[Conceptual illustration\label{fig:causal_anticausal_illustration}]{\includegraphics[width=0.415\linewidth]{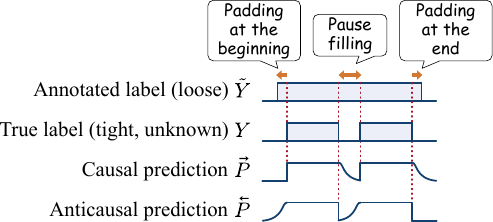}}
\hfill
\subfloat[Labels and actual inference results for three 10-second chunks\label{fig:causal_anticausal_real}]{\includegraphics[width=0.55\linewidth]{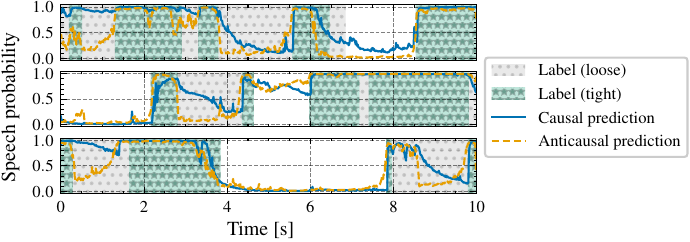}}
\vspace{-0.7em}
\caption{Behavior of causal and anticausal models in single-speaker cases.}
\end{figure*}

\section{Problem formulation}
\subsection{Formulation of the conventional speaker diarization}
Speaker diarization is the problem of estimating speaker-wise speech activity at each frame.
Let $X=\left[\vect{x}_1,\dots,\vect{x}_T\right]\in\mathbb{R}^{D\times T}$ denote frame-wise $D$-dimensional acoustic features, where $T$ is the number of frames.
Given $X$, the goal is to estimate the speaker activities $\hat{Y}=\left[\vect{\hat{y}}_1,\dots,\vect{\hat{y}}_{T}\right]\in\{0,1\}^{S\times T}$, where $S$ is the number of speakers and $\vect{\hat{y}}_t\coloneqq\trans{\left[\hat{y}_{1,t},\dots,\hat{y}_{S,t}\right]}$ represents the speaker activity vector, with $\hat{y}_{s,t}=1$ indicating that speaker $s$ is active at time frame $t$.

In the EEND framework~\cite{fujita2019end1,fujita2019end2}, a model is trained to output frame-wise posterior probabilities $P=\left[\vect{p}_1,\dots,\vect{p}_T\right]\in\left(0,1\right)^{S\times T}$ where $\vect{p}_t\coloneqq\trans{\left[p_{1,t},\dots,p_{S,t}\right]}$, which yield $\hat{Y}$ by applying a threshold $\tau$:
\begin{equation}
    \hat{y}_{s,t}=\begin{cases}
    1 & (p_{s,t}\geq\tau),\\
    0 & (p_{s,t}<\tau).
    \end{cases}
\end{equation}
Hereafter, we use $\phi_{\tau}(\cdot)$ to denote the elementwise extension of the above thresholding rule, yielding $\hat{Y}=\operatorname{\phi_{\tau}}(P)$, and we set $\tau=0.5$ throughout this paper.
Such a model is trained in a supervised manner using the loss between the true label $Y=(y_{s,t})\in\left\{0,1\right\}^{S\times T}$ and the prediction $P$.
Since the model output is permutation-free in the speaker order, it is necessary to first find speaker correspondence between them.
Let $P^{\left(Y\right)}\coloneqq\Pi_{P,Y}P$ denote the permutation of $P$ whose speaker order best matches $Y$, where the permutation matrix $\Pi_{P,Y}$ is defined as
\begin{equation}
\Pi_{P,Y}\coloneqq\argmin_{\Pi\in\mathbb{P}_S}\norm{\Pi P- Y}_F.\label{eq:permutation}
\end{equation}
Here, $\mathbb{P}_S$ is the set of $S\times S$ permutation matrices and $\norm{\cdot}_F$ denotes the Frobenius norm.

In this paper, we adopt a variant of EEND that outputs posterior probabilities for each powerset class of speakers~\cite{plaquet2023powerset}.
Let $\operatorname{\mathcal{P}}\left(S\right)$ denote the powerset of the speaker index set $\left\{1,\dots,S\right\}$.
In this case, the model outputs $\vect{q}_{t}\coloneqq\trans{\left[q_{1,t}\dots,q_{R,t}\right]}\in\left(0,1\right)^R$ at each frame $t$, where $q_{r,t}$ represents the probability of the $r$-th class corresponding to the speaker set $\mathcal{R}_r\in\operatorname{\mathcal{P}}\left(S\right)$.
Following the previous studies~\cite{plaquet2023powerset}, we only consider the overlap of at most two speakers, and thus restrict $\mathcal{R}_r$ such that $\abs{\mathcal{R}_r}\leq 2$, which results in $R=\sum_{i=0}^2\binom{S}{i}$.
In this paper, we set $S=4$, so the set of possible classes is given by $\mathcal{R}_r\in\{\emptyset,\{1\},\allowbreak\{2\},\allowbreak\{3\},\allowbreak\{4\},\allowbreak\{1,2\},\allowbreak\{1,3\},\allowbreak\{1,4\},\allowbreak\{2,3\},\allowbreak\{2,4\},\allowbreak\{3,4\}\}$.
We define a mapping $\mathsf{Q2P}:Q\mapsto P$ from the powerset posterior $Q=\left[\vect{q}_1,\dots,\vect{q}_T\right]$ to the speaker-wise posterior $P$, where the conversion is given by
\begin{equation}
    p_{s,t}=\sum_{r:\,s\in\mathcal{R}_r} q_{r,t}.\label{eq:q_to_p}
\end{equation}
Using the posterior probabilities for the powerset classes, the training objective to be minimized is defined as the multi-class cross-entropy loss:
\begin{equation}
\operatorname{\ell}\bigl(Q^{\left(Y\right)},Y\bigr)=-\frac{1}{T}\sum_{t=1}^T \sum_{\substack{\mathcal{R}_r\in\operatorname{\mathcal{P}}(S)\\\abs{\mathcal{R}_r}\leq 2}}y'_{r,t}\log q^{(Y)}_{r,t},\label{eq:loss}
\end{equation}
where $y'_{r,t}=1$ if the reference label $\vect{y}_{t}$ corresponds to $\mathcal{R}_r$, and $y'_{r,t}=0$ otherwise, and $Q^{(Y)}=(q^{(Y)}_{r,t})$ is the permuted version of $Q$, aligned to $Y$ using the corresponding $P$ derived from \autoref{eq:q_to_p}.

\subsection{Formulation of speaker diarization in this study: Treating annotated labels as weak supervision}
Prior studies treated annotated label $\tilde{Y}=\left(\tilde{y}_{s,t}\right)\in\left\{0,1\right\}^{S\times T}$ as the ground truth during training, i.e., $Y=\tilde{Y}$.
However, as noted earlier, the labeling criteria differ substantially between ASR and diarization corpora.
In contrast, we consistently regard tight labels as the true labels $Y$.
We assume that the annotated labels in ASR corpora have loose boundaries, i.e., frames annotated as speech may in fact be silent:\footnote{In practice, the speech intervals obtained via forced alignment may fall outside the loose annotation. However, the extent of such deviations is very small (less than \SI{2}{\percent} according to \cite{horiguchi2025can}). Therefore, in this work, we define the problem under the assumption that the tight intervals are always contained within the loose intervals.}
\begin{align}\label{eq:label}
y_{s,t}\in\begin{cases}
\left\{0,1\right\} & (\tilde{y}_{s,t}=1),\\
\left\{0\right\} & (\tilde{y}_{s,t}=0).
\end{cases}
\end{align}
This paper addresses the problem of training a model to estimate the unknown tight labels $Y$ using only the annotated, potentially loose labels $\tilde{Y}$.

In this paper, we train models using a compound set of ASR and diarization corpora.
We assume that it is known a priori whether each sample originates from an ASR corpus (i.e., loosely annotated) or a diarization corpus (i.e., tightly annotated).
For ASR corpora, we generate pseudo labels by tightening the original annotations $\tilde{Y}$ using the proposed methods described in the following section.
For diarization corpora, we assume that $Y=\tilde{Y}$.

\section{Proposed method}
\subsection{Concept}
As illustrated in \autoref{fig:task}, conventional training leads speaker diarization models to estimate not only who is speaking when but also to pad segment boundaries and fill pauses.
If the former functionality can be isolated, it would benefit downstream tasks.
However, once these functions are internalized in a diarization model, disentangling them becomes challenging.

Considering how the latter functionality is realized, it likely requires the model to first identify latent tight speech segments and then use contextual information to produce loose boundaries that match the looseness of the so-called ground truth labels.
Leveraging this insight, we propose a method that deliberately employs causal and anticausal architectures to prevent the model from performing padding segment boundaries and filling pauses, thereby separating these functionalities.
\footnote{Interested readers may also refer to a study on sound event detection using causal and anticausal models with weak labels~\cite{ebbers2020forward}.}

We first explain the operating principle of the proposed method in a simple single-speaker case in \autoref{sec:single}.
Next, in \autoref{sec:multi}, we discuss additional challenges that arise in the multi-speaker setting, typically addressed by speaker diarization, along with their solutions.
Finally, in \autoref{sec:cotraining}, we introduce a training strategy that enables more efficient use of the proposed method.

\subsection{Preliminary: Label tightening in single-speaker cases}\label{sec:single}
\subsubsection{Causal--anticausal consistency}
First, for simplicity, we consider single-speaker cases, i.e., $S=1$.
To determine how much to pad segment boundaries or how long to fill a silent interval, a model must first identify tightly bounded speech segments and then examine the surrounding context.
Since diarization models are typically based on bidirectional~\cite{fujita2019end1}, fully attentive~\cite{fujita2019end2,horiguchi2022encoderdecoder}, or large-receptive-field architectures~\cite{zeghidour2021dive,horiguchi2025pretraining}, such capabilities tend to be acquired implicitly.
However, disentangling the detection of tight speech segments from context-based segment expansion and pause filling is difficult because all of these functions are learned together.
In this paper, we intentionally adopt a unidirectional architecture that deprives the model of such abilities, with the aim of separating these functions.
Specifically, we employ two unidirectional models: one can only see the past frames, and the other can only see the future frames.
For simplicity, we call the former the \textit{causal model} and the latter the \textit{anticausal model}.

\begin{figure*}[t]
\begin{minipage}[t]{0.22\linewidth}
\vspace{0pt}
\subfloat[Labels and predictions\label{fig:definition}]{\includegraphics[width=\linewidth]{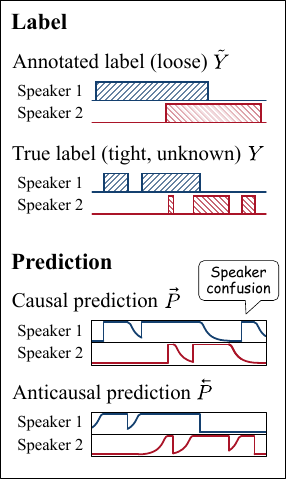}}
\end{minipage}
\hfill
\begin{minipage}[t]{0.76\linewidth}
\vspace{0pt}
\subfloat[Basic tightening\label{fig:basic_tightening}]{\includegraphics[width=\linewidth]{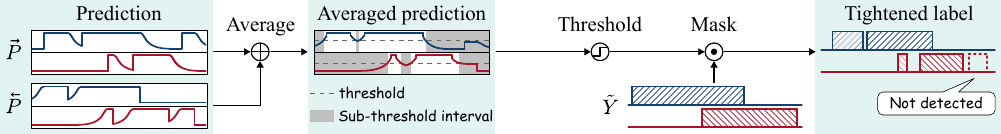}}\vspace{-0.5em}\\
\subfloat[VAD tightening\label{fig:vad_tightening}]{\includegraphics[width=\linewidth]{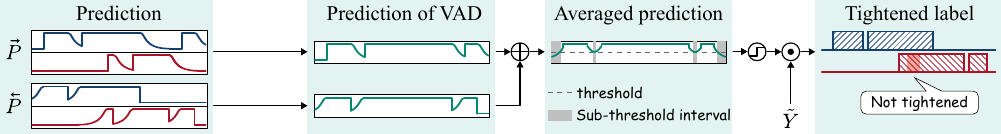}}\vspace{-0.5em}\\
\subfloat[SC tightening\label{fig:sc_tightening}]{\includegraphics[width=\linewidth]{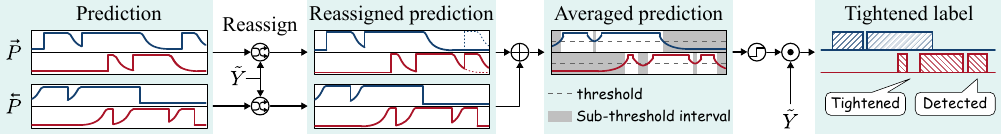}}
\end{minipage}
\vspace{-0.7em}
\caption{Example of label tightening in a two-speaker case.}\label{fig:tightening}
\end{figure*}

We first describe the behavior of the causal model (\autoref{fig:causal_anticausal_illustration}, third row).
Since it cannot access future frames, it cannot pad the beginning of a speech segment, even when trained with loose labels.
Moreover, once speech is interrupted, the model cannot predict whether it will resume shortly afterward.
As a result, it cannot determine whether a subsequent silent interval should be treated as part of a speech segment or as the end of speech.
Consequently, the model becomes progressively less confident after speech offset, resulting in lower posterior probabilities.
The anticausal model exhibits analogous behavior.
It cannot pad the end of a speech segment, and the posterior probability before speech onset gradually increases, as shown in the fourth row of \autoref{fig:causal_anticausal_illustration}.
Representative examples from actual predictions are also shown in \autoref{fig:causal_anticausal_real}.

From these observations, both the causal and anticausal models should yield high speech posteriors over the true speech intervals.
In contrast, in the padded or filled regions, one of the two models is expected to fail to detect speech in principle.
Therefore, by appropriately combining the outputs of these two models, it becomes possible to recover tight speech segments.
In this paper, we simply take the average of posteriors followed by thresholding for combination.

It is also possible to use this thresholded average directly as the tightened label; however, because it is based on inference, it inevitably contains inference errors.
Among these errors, missed detections cannot be distinguished from those caused by the tightening process itself.
In contrast, false-alarm speech can be filtered out by exploiting the property in \autoref{eq:label}, namely that regions annotated as silence in the loose labels are guaranteed to be silent.
This filtering is equivalent to masking the loose speech segments using the causal prediction $\causal{P}\in(0,1)^{S\times T}$ and anticausal prediction $\anticausal{P}\in(0,1)^{S\times T}$.
The final tightened labels are given by
\begin{equation}
    Y_\mathrm{tight}=\tilde{Y}\odot\operatorname{\phi_\tau}\biggl(\frac{\causal{P}+\anticausal{P}}{2}\biggr),\label{eq:tightening}
\end{equation}
where $\odot$ denotes the Hadamard product.
This $Y_\mathrm{tight}$ is used as the target labels for training the final non-causal model.

\subsubsection{Restoration of over-tightened speech segments}\label{sec:restore}
It is difficult to determine whether the reduction of speech from $\hat{Y}$ to $Y_\mathrm{tight}$ results from a correct tightening or from erroneous deletions caused by inference errors.
Moreover, when this label update is repeatedly applied, as described in \autoref{sec:cotraining}, the method may suffer from over-tightening.
Although annotated segment boundaries may be inaccurate, each annotated interval should still contain speech and thus should not be removed entirely.
We thus propose restoring lost segments due to tightening using the corresponding segments from the loose annotations.
In this paper, we relax the constraint by regarding cases in which more than \SI{50}{\percent} of a segment is removed as excessive deletion, and restore the corresponding loose segment.

\subsection{Label tightening in multi-speaker cases}\label{sec:multi}
For speaker diarization purposes, the tightening method described in \autoref{sec:single} should be extended to multi-speaker cases.
We introduce the following three types of tightening methods.

\subsubsection{Basic tightening}
The simplest variant applies the procedure in \autoref{sec:single} speaker-wise, as depicted in \autoref{fig:basic_tightening}.
Since the orders of speakers in $\causal{P}$, $\anticausal{P}$, and $\tilde{Y}$ are not necessarily the same, we first align $\causal{P}$ and $\anticausal{P}$ and then align them with $\tilde{Y}$ by
\begin{align}
\anticausal{P}&\leftarrow\Pi_{\anticausal{P},\causal{P}}\anticausal{P},\label{eq:align_causal_anticausal}\\
\causal{P},\anticausal{P}&\leftarrow\Pi_{\frac{\causal{P}+\anticausal{P}}{2},\tilde{Y}}\causal{P},\Pi_{\frac{\causal{P}+\anticausal{P}}{2},\tilde{Y}}\anticausal{P}.\label{eq:align_causal_label}
\end{align}
We then apply the tightening procedure in \autoref{sec:single} in a speaker-wise manner.
To avoid redundancy, the following explanation of the tightening methods assumes that the speaker order has already been aligned by \autoref{eq:align_causal_anticausal}--\autoref{eq:align_causal_label}.

\subsubsection{VAD tightening}\label{sec:vad_tightening}
In multi-speaker cases, inference results may include speaker confusion additionally.
In the example shown in \autoref{fig:tightening}, we assume that the final speech segment of the speaker 2 is incorrectly predicted as the speaker 1's speech in the output of the causal model (\autoref{fig:definition}).
In such a situation, the same speech segment would be attributed to different speakers in the causal and anticausal outputs.
When the posteriors are averaged as in the basic tightening approach, the label-tightening mask may disappear for both speakers (\autoref{fig:basic_tightening}).

To address this issue, the second tightening method is designed to be robust against speaker confusion by constructing the mask based on voice activity detection (VAD), as shown in \autoref{fig:vad_tightening}.
Because the models output probabilities over the powerset classes, the silence-class output ($\causal{q}_{0,t}$ and $\anticausal{q}_{0,t}$ for causal and anticausal output, respectively) can be regarded as the complement of the speech probabilities.
Let $\causal{Q}_{>0}\in(0,1)^{S\times T}$ be obtained by stacking $S$ copies of $\left[1-\causal{q}_{0,1},\dots,1-\causal{q}_{0,T}\right]$ vertically, and define $\anticausal{Q}_{>0}$ analogously.
Using these, we obtain the tightened labels as follows, replacing \autoref{eq:tightening}:
\begin{equation}
Y_\mathrm{tight}=\tilde{Y}\odot\phi_\tau\biggl(\frac{\causal{Q}_{>0}+\anticausal{Q}_{>0}}{2}\biggr).
\end{equation}
This corresponds to removing regions predicted as non-speech from the annotations.

\subsubsection{SC tightening}
While VAD tightening mitigates speaker confusion in predictions, it becomes overly conservative and is ineffective in overlapped intervals, as shown in \autoref{fig:vad_tightening}.
In the third approach, we trust the number of speakers estimated at each frame, whereas VAD tightening relies solely on speech presence.
We refer to this method as speaker-counting (SC) tightening.

\begin{algorithm}[t]
    \caption{Posterior reassignment in SC tightening}\label{algo:sc_tightening}
    \DontPrintSemicolon
    \SetKwInOut{Input}{Input}
    \SetKwInOut{Output}{Output}
    \SetKwComment{Comment}{$\triangleright$\ }{}
    \Input{Prediction $P=\{p_{s,t}\}$, annotated label $\tilde{Y}=\{\tilde{y}_{s,t}\}$, threshold $\tau$}
    \Output{Reassigned prediction $P$}
    \BlankLine
    \For{$t=1$~\KwTo~$T$} {
        $\mathcal{M}\leftarrow\emptyset$\Comment*[r]{Indices of missed speakers}
        $\mathcal{F}\leftarrow\emptyset$\Comment*[r]{Indices of false alarm}
        \For(\Comment*[f]{Find candidates}){$s=1$~\KwTo~$S$\label{algline:find_start}}{
            \uIf(\Comment*[f]{Missed speaker}){$\tilde{y}_{s,t}=1 \land p_{s,t}\le\tau$}{
                $\mathtt{PushBack}(\mathcal{M},s)$
            }
            \ElseIf(\Comment*[f]{False alarm}){$\tilde{y}_{s,t}=0 \land p_{s,t}>\tau$}{
                $\mathtt{PushBack}(\mathcal{F},s)$\label{algline:find_end}
            }
        }
        \While(\Comment*[f]{Reassignment}){$\mathcal{M}\neq\emptyset \land \mathcal{F}\neq\emptyset$\label{algline:swap_start}}{
            $s_1\leftarrow\mathtt{PopFront}(\mathcal{M})$\;
            $s_2\leftarrow\mathtt{PopFront}(\mathcal{F})$\;
            $\mathtt{Swap}(p_{s_1,t},p_{s_2,t})$\;\label{algline:swap_end}
        }
    }
\end{algorithm}

The SC tightening procedure is summarized in \autoref{algo:sc_tightening}.
In short, at each time frame, it identifies the missed speaker(s) and the false-alarmed speaker(s) (\autoref{algline:find_start}--\ref{algline:find_end}), and swaps their posterior probabilities as much as possible (\autoref{algline:swap_start}--\ref{algline:swap_end}).
This enables label tightening while mitigating the impact of speaker confusion in the predictions, as illustrated in \autoref{fig:sc_tightening}.

It may be noted that when multiple missed or false-alarm speakers require posterior swapping, the reassignment process depends on the ordering of speakers in $P$ and $\tilde{Y}$.
This is because the swapping procedure terminates when either list becomes empty (\autoref{algline:swap_start}).
To address this, we reorder the speakers in $\tilde{Y}$ in advance in descending order of speaking duration.
This prioritizes speakers for whom speaker confusion would have a larger impact if left uncorrected and makes the algorithm deterministic.
Note that the speaker ordering in $P$ is already permuted to align with $\tilde{Y}$ based on the correlation of their speech intervals via \autoref{eq:permutation}.
Therefore, cases in which multiple speakers simultaneously become swap candidates are rare in practice, given the limited number of speakers ($S=4$).

\subsection{Progressive label tightening via co-training of causal and anticausal models}\label{sec:cotraining}
The methods described in \autoref{sec:multi} enable us to obtain labels that are tighter than the original annotations.
However, since both the causal and anticausal models are themselves trained with loose labels, they tend to reproduce the looseness of the annotations as much as possible.
As a result, the tightening effect remains limited.
A straightforward approach to overcome this issue is to retrain the causal and anticausal models with the tightened labels and repeat the tightening process based on their outputs.
Repeating this procedure is expected to progressively produce tighter labels.
In practice, however, this approach is inefficient because each loop requires separate training of the causal and anticausal models, while the improvement achieved within a single loop is marginal.

\begin{algorithm}[t]
    \caption{Co-training of causal and anticausal models}\label{algo:cotraining}
    \DontPrintSemicolon
    \SetKwInOut{Input}{Input}
    \SetKwInOut{Output}{Output}
    \SetKwComment{Comment}{$\triangleright$\ }{}
    \Input{Training dataset $\mathcal{D}=\{(X_i,\tilde{Y}_i)\}_{i=1}^N$, causal model $\causal{f}$ with parameters $\causal{\theta}$, anticausal model $\anticausal{f}$ with parameters $\anticausal{\theta}$}
    \Output{Updated parameters $\causal{\theta}$ and $\anticausal{\theta}$}
    \BlankLine
    \ForEach{minibatch $(\mathbf{X},\tilde{\mathbf{Y}})$}{
        $\causal{\mathbf{Q}}\leftarrow \causal{f}(\mathbf{X};\causal{\theta})$\Comment*[r]{Forward causal}\label{algline:forward_causal}
        $\anticausal{\mathbf{Q}}\leftarrow\anticausal{f}(\mathbf{X};\anticausal{\theta})$\Comment*[r]{Forward anticausal}\label{algline:forward_anticausal}
        $\causal{\mathbf{P}},\anticausal{\mathbf{P}}\leftarrow\mathsf{Q2P}(\causal{\mathbf{Q}}),\mathsf{Q2P}(\anticausal{\mathbf{Q}})$\Comment*[r]{Eq. \autoref{eq:q_to_p}}\label{algline:conversion}
        $\mathbf{Y}_\mathrm{tight}\leftarrow \mathsf{Tighten}\bigl(\tilde{\mathbf{Y}},\causal{\mathbf{P}},\anticausal{\mathbf{P}}\bigr)$\Comment*[r]{\autoref{sec:multi}}\label{algline:tighten}
        $\mathbf{Y}_\mathrm{tight}\leftarrow \mathsf{Restore}\bigl(\mathbf{Y}_\mathrm{tight},\tilde{\mathbf{Y}}\bigr)$\Comment*[r]{\autoref{sec:restore}}\label{algline:restore}
        Update $\causal{\theta}$ on $\ell(\causal{\mathbf{Q}},\mathbf{Y}_\mathrm{tight})$\label{algline:backward_causal}\Comment*[r]{Backward causal}
        Update $\anticausal{\theta}$ on $\ell(\anticausal{\mathbf{Q}},\mathbf{Y}_\mathrm{tight})$ \label{algline:backward_anticausal}\Comment*[r]{Backward anticausal}
    }
\end{algorithm}

To resolve this inefficiency, we propose the co-training of causal and anticausal models, in which tightening is applied at every minibatch.
The pseudocode is provided in \autoref{algo:cotraining}.
We assume that both causal and anticausal models have been trained using loose labels, resulting in parameters $\causal{\theta}$ and $\anticausal{\theta}$, respectively.
Note that bold symbols denote batched tensors, e.g., $\mathbf{X}\in\mathbb{R}^{B\times D\times T}$, where $B$ is the batch size.
The loss function defined on a single sample in \autoref{eq:loss} and the posterior conversion $\mathsf{Q2P}$ are extended to batched inputs by applying them independently along the batch dimension.
In each training step, we first run forward passes of both the causal and anticausal models to obtain their inference results (\autoref{algline:forward_causal}--\ref{algline:forward_anticausal}), followed by powerset-to-multilabel conversion (\autoref{algline:conversion}).
We then apply tightening and restoration to generate refined labels for the minibatch (\autoref{algline:tighten}--\ref{algline:restore}).
Finally, the tightened labels are used as supervision to compute the loss, and the parameters of both models are updated accordingly (\autoref{algline:backward_causal}--\ref{algline:backward_anticausal}).

Once co-training is complete, we infer all ASR-corpus samples in the training dataset using the causal and anticausal models with the resulting parameters $\causal{\theta}$ and $\anticausal{\theta}$.
We then train a final non-causal model from scratch using the tightened labels as supervision.

\section{Experimental settings}
\subsection{Speaker diarization pipeline}
We adopted the EEND-vector clustering framework~\cite{kinoshita2021integrating}, which consists of i) local diarization with a 10-second window and a 1-second shift, ii) speaker embedding extraction for each detected speaker in each window, and iii) clustering of the speaker embeddings to determine speaker correspondence across windows.
The proposed method is applied to the local diarization models, whose detailed configurations are described in \autoref{sec:model}.
For speaker embedding extraction, we use ECAPA-TDNN trained on the VoxCeleb 1\&2 datasets~\cite{nagrani2017voxceleb}.
For clustering, we use agglomerative hierarchical clustering.
The clustering parameters, i.e., the threshold value and the minimum cluster size, were tuned using the validation set of the compound dataset.

\subsection{Local diarization model}\label{sec:model}
Many high-performing diarization models rely on large-scale self-supervised pretrained models~\cite{han2025leveraging,han2025fine,han2025efficient,plaquet2025mamba}.
A representative example is WavLM~\cite{chen2022wavlm}, which is publicly available.
However, it does not support causal and anticausal processing, and training such models from scratch is not feasible for everyone.
For proof of concept, we instead use models based on a speaker embedding extractor pretrained via multi-speaker identification~\cite{horiguchi2025pretraining}, which offers a good trade-off between performance and training cost.
Similarly, we use ECAPA-TDNN~\cite{desplanques2020ecapatdnn} and ReDimNet~\cite{yakovlev2024reshape} as the frontend encoders, respectively, each followed by a classification backend.
The specific pretraining procedure using the VoxCeleb 1\&2 datasets~\cite{nagrani2020voxceleb} also follows that described in the reference paper.

The causal and anticausal models are implemented by replacing each component in the architectures with a unidirectional version.
Our purpose here is not to implement perfectly streamable models, but rather to prevent the models from padding segment boundaries or filling pauses based on contextual information.
Therefore, some components are intentionally left unchanged, even though they are not streaming-ready.
Detailed implementation is described in the following subsections.

In this section, we restrict our discussion to causal models, where only past frames are accessible at each time step.
The anticausal case is analogous, with access limited to future frames.

\subsubsection{ECAPA-TDNN}
The non-causal parts of the original ECAPA-TDNN are 1D convolutional layers, batch normalization, and squeeze-and-excitation (SE) blocks~\cite{desplanques2020ecapatdnn}.

\noindent\textbf{1D convolutional layers.}
We simply replaced each with the causal alternative.
Their kernel size was kept unchanged.

\noindent\textbf{Batch normalization.}
While batch normalization also incorporates future information during training, inference relies solely on the learned mean and variance. As a result, no future inputs are accessed during inference, and operational causality is maintained. Hence, batch normalization is retained without modification.

\noindent\textbf{SE blocks.}
SE blocks recalibrate channel-wise importance based on the global average~\cite{hu2018squeeze}.
Let $Z\coloneqq\left[\vect{z}_1,\dots,\vect{z}_T\right]\in\mathbb{R}^{D\times T}$ be an input sequence of vector representations and $\bar{\vect{z}}\coloneqq\frac{1}{T}\sum_{t=1}^T\vect{z}_t$.
The operations in the SE block are denoted as
\begin{align}
    \vect{s}&=g(\vect{\bar{z}})\in\left(0,1\right)^{D},\\
    Z'&=Z\odot[\underbrace{\vect{s},\dots,\vect{s}}_{T}]\in\mathbb{R}^{D\times T},
\end{align}
where $g(\cdot)$ is a nonlinear transformation implemented by two-stacked fully connected layers.
The SE block is not causal because it relies on the global average $\bar{\vect{z}}$.
While its causal variant has been proposed~\cite{yu2021dual}, we found that it led to performance degradation.
Simply leveraging a global average does not enable local context-dependent boundary padding or pause filling, and therefore does not conflict with the objective of this study.
Accordingly, we retrain the original SE block without modification in the causal model.

\subsubsection{ReDimNet}
We use the B2 variant of ReDimNet~\cite{yakovlev2024reshape}.
The non-causal part of ReDimNet is 1D/2D convolutional layers, Transformer encoders, and batch normalization.

\noindent\textbf{1D/2D convolutional layers.}
As in ECAPA-TDNN, both layers were replaced with causal versions using the same kernel sizes.

\noindent\textbf{Transformer encoders.}
For each self-attention module in Transformer encoders, we applied a causal mask that prevents each query from attending to future keys.

\noindent\textbf{Batch normalization.}
As in ECAPA-TDNN, we did not replace batch normalization with a causal variant.

\subsubsection{Classification backend}
The classification backend consists of a single-layer long short-term memory (LSTM) network followed by a fully-connected layer, following the prior study~\cite{horiguchi2025pretraining}.
For the causal model, we use a unidirectional LSTM in the left-to-right direction.

\subsection{Diarization dataset}
\begin{table}[t]
\caption{Components of the compound diarization dataset.}
\label{tbl:dataset}
\centering
\setlength{\tabcolsep}{3pt}
\resizebox{\linewidth}{!}{%
\begin{tabular}{@{}llrrr@{}}
\toprule
&&\multicolumn{3}{c@{}}{Duration (hours)}\\\cmidrule(l{\tabcolsep}){3-5}
Corpus & Label &Train & Val & Test\\\midrule
AMI-MHM~\cite{carletta2007unleashing} &ASR-oriented (loose)&81&10&9\\
AMI-SDM~\cite{carletta2007unleashing} & ASR-oriented (loose) &80 & 10 & 9\\
AliMeeting~\cite{yu2022m2met} & ASR-oriented (loose)&111 & 4 & 11\\
MSDWild~\cite{liu2022msdwild} &Diarization-oriented (tight)&64 & 2 & 10\\
VoxConverse~\cite{chung2020spot} & Diarization-oriented (tight)& 18 & 2 & 44\\\midrule
Total&&354&28&82\\
\bottomrule
\end{tabular}%
}
\end{table}

We use the compound dataset of ASR and diarization corpora in our experiments, which is standard in recent speaker diarization studies~\cite{bredin2023pyannote,plaquet2023powerset,horiguchi2025pretraining,han2025efficient}.
While the proposed method can progressively tighten loose annotations without tight supervision, evaluating the validity of the resulting labels still requires ideal tight labels.
In this paper, for ASR-oriented data, we use the corpora for which forced-alignment-based labels are publicly available~\cite{horiguchi2025can}\footnote{\url{https://github.com/nttcslab-sp/diar-forced-alignment}}: AMI mixed headset microphones (AMI-MHM)~\cite{carletta2007unleashing}, AMI single distant microphone (AMI-SDM)~\cite{carletta2007unleashing}, and AliMeeting~\cite{yu2022m2met}.
As the diarization corpora, we use the MSDWild few-talker set (MSD)~\cite{liu2022msdwild} and VoxConverse (VC)~\cite{chung2020spot}.
The sizes of each corpus and the compound dataset are listed in \autoref{tbl:dataset}.
We also use DIHARD III~\cite{ryant2021third} as an out-of-domain corpus exclusively for evaluation, which is tightly annotated with the original labels.

\begin{table*}[!t]
\caption{DERs (\%) on the in-domain corpora. For DER computation, we used tight labels obtained via forced alignment as the reference. For the ASR corpora, the relative improvement achieved by the proposed method is shown in parentheses, where the DER with loose-label training is treated as the baseline and that with tight-label training as the topline.}
\renewrobustcmd{\bfseries}{\fontseries{b}\selectfont}
\renewrobustcmd{\boldmath}{}
\newrobustcmd{\B}{\bfseries}
\label{tbl:results_indomain}
\centering
\sisetup{detect-weight,mode=text}
\setlength{\tabcolsep}{3pt}
\resizebox{\linewidth}{!}{%
\begin{tabular}{@{}lcc*{3}{S[table-format=2.2]@{\,}r}*{2}{S[table-format=2.2]}*{3}{S[table-format=2.2]@{\,}r}*{2}{S[table-format=2.2]}@{}}
\toprule
&\multicolumn{2}{c}{\multirow{2.92}{*}{\makecell{Supervision for\\ASR corpora}}}&\multicolumn{8}{c}{Encoder: ECAPA-TDNN}&\multicolumn{8}{c}{Encoder: ReDimNet}\\\cmidrule(l{\tabcolsep}r{\tabcolsep}){4-11}\cmidrule(l{\tabcolsep}){12-19}
&&&\multicolumn{6}{c}{Corpus type: ASR}&\multicolumn{2}{c}{Diarization}&\multicolumn{6}{c}{Corpus type: ASR}&\multicolumn{2}{c@{}}{Diarization}\\\cmidrule(l{\tabcolsep}r{\tabcolsep}){2-3}\cmidrule(l{\tabcolsep}r{\tabcolsep}){4-9}\cmidrule(l{\tabcolsep}r{\tabcolsep}){10-11}\cmidrule(l{\tabcolsep}r{\tabcolsep}){12-17}\cmidrule(l{\tabcolsep}){18-19}
Model&Label&Tightening&\multicolumn{2}{c}{AMI-MHM}&\multicolumn{2}{c}{AMI-SDM}&\multicolumn{2}{c}{AliMeeting}&\multicolumn{1}{c}{MSD}&\multicolumn{1}{c}{VC}&\multicolumn{2}{c}{AMI-MHM}&\multicolumn{2}{c}{AMI-SDM}&\multicolumn{2}{c}{AliMeeting}&\multicolumn{1}{c}{MSD}&\multicolumn{1}{c@{}}{VC}\\\midrule
\multicolumn{11}{@{}l}{\textbf{Baseline \& topline}}\\
($\mathtt{B1}$) Non-causal&Loose&N/A&35.24&{\scriptsize (\SI{0.0}{\percent})}&38.79&{\scriptsize (\SI{0.0}{\percent})}&29.39&{\scriptsize (\SI{0.0}{\percent})}&27.53&12.32&34.92&{\scriptsize (\SI{0.0}{\percent})}&38.51&{\scriptsize (\SI{0.0}{\percent})}&27.67&{\scriptsize (\SI{0.0}{\percent})}&26.07&11.38\\
($\mathtt{B2}$) Non-causal& Tight&N/A&16.12&{\scriptsize (\SI{100.0}{\percent})}&20.54&{\scriptsize (\SI{100.0}{\percent})}&23.49&{\scriptsize (\SI{100.0}{\percent})}&26.23&11.75&14.83&{\scriptsize (\SI{100.0}{\percent})}&19.48&{\scriptsize (\SI{100.0}{\percent})}&20.64&{\scriptsize (\SI{100.0}{\percent})}&26.19&12.09\\
\midrule
\multicolumn{11}{@{}l}{\textbf{Proposed method}}\\
($\mathtt{P1}$) Non-causal&  Loose & Basic&23.36&{\scriptsize (\SI{62.1}{\percent})}&25.94&{\scriptsize (\SI{70.4}{\percent})}&26.93&{\scriptsize (\SI{41.7}{\percent})}&27.52&12.26&20.37&{\scriptsize (\SI{72.4}{\percent})}&25.37&{\scriptsize (\SI{69.0}{\percent})}&24.81&{\scriptsize (\SI{40.7}{\percent})}&25.55&11.02\\
($\mathtt{P2}$) Non-causal&Loose& VAD&24.30&{\scriptsize (\SI{57.2}{\percent})}&28.29&{\scriptsize (\SI{57.5}{\percent})}&26.40&{\scriptsize (\SI{50.7}{\percent})}&28.09&12.21&19.73&{\scriptsize (\SI{75.6}{\percent})}&25.16&{\scriptsize (\SI{70.2}{\percent})}&\B 23.69&{\scriptsize (\SI{56.6}{\percent})}&27.08&11.96\\
($\mathtt{P3}$) Non-causal&Loose&SC&\B 20.96&{\scriptsize (\SI{74.7}{\percent})}&\B 25.29&{\scriptsize (\SI{74.0}{\percent})}&\B 26.10&{\scriptsize (\SI{55.8}{\percent})}& 27.28&12.29&\B 19.14&{\scriptsize (\SI{78.5}{\percent})}&\B 23.62&{\scriptsize (\SI{78.2}{\percent})}&23.75&{\scriptsize (\SI{55.8}{\percent})}&26.08&11.56\\
\bottomrule
\end{tabular}%
}\\
\end{table*}

\begin{table}[t]
\caption{DER (\%) on DIHARD III, an out-of-domain corpus annotated with diarization-oriented tight labels.}\label{tbl:results_dihard}
\renewrobustcmd{\bfseries}{\fontseries{b}\selectfont}
\renewrobustcmd{\boldmath}{}
\newrobustcmd{\B}{\bfseries}
\resizebox{\linewidth}{!}{%
\begin{tabular}{@{}lcc*{2}{S[table-format=2.2]@{\,}r}@{}}
\toprule
&\multicolumn{2}{c}{Supervision for}\\
&\multicolumn{2}{c}{ASR corpora}&\multicolumn{4}{c@{}}{Encoder}\\\cmidrule(l{\tabcolsep}r{\tabcolsep}){2-3}\cmidrule(l{\tabcolsep}){4-7}
Model&Label&Tightening&\multicolumn{2}{c}{ECAPA-TDNN}&\multicolumn{2}{c@{}}{ReDimNet}\\\midrule
\multicolumn{5}{@{}l}{\textbf{Baseline \& topline}}\\
($\mathtt{B1}$) Non-causal&Loose&N/A&29.96&{\scriptsize (\SI{0.0}{\percent})}&29.89&{\scriptsize (\SI{0.0}{\percent})}\\
($\mathtt{B2}$) Non-causal&Tight&N/A&25.63&{\scriptsize (\SI{100.0}{\percent})}&26.07&{\scriptsize (\SI{100.0}{\percent})}\\\midrule
\multicolumn{5}{@{}l}{\textbf{Proposed method}}\\
($\mathtt{P1}$) Non-causal&Loose&Basic&27.29&{\scriptsize (\SI{61.7}{\percent})}&26.52&{\scriptsize (\SI{88.2}{\percent})}\\
($\mathtt{P2}$) Non-causal&Loose&VAD&28.38&{\scriptsize (\SI{36.5}{\percent})}&26.38&{\scriptsize (\SI{91.9}{\percent})}\\
($\mathtt{P3}$) Non-causal&Loose&SC&\B 27.06&{\scriptsize (\SI{67.0}{\percent})}&\B 25.28&{\scriptsize (\SI{120.7}{\percent})}\\\bottomrule
\end{tabular}%
}
\end{table}

\subsection{Training of diarization models}
Except for the co-training stage, the frontend encoders were initialized with weights obtained from multi-speaker identification pretraining, while the classification backend was trained from random weights.
The batch size was set to 32, and training was conducted for up to 180k steps.
We used the Adam optimizer~\cite{kingma2015adam}, with the learning rate linearly increased to 0.001 over the first 1k steps, followed by exponential decay with a factor of 0.8 every 6k steps.
In co-training, we initialize the causal and anticausal models with the weights obtained from the above training and further train them for up to 6k steps using the procedure described in \autoref{algo:cotraining}.
The optimization is performed using Adam with a fixed learning rate of 0.0001.

In co-training and subsequent training with tightened labels, it is difficult to properly assess convergence unless the validation set has tightly annotated labels.
In our experiments, we assume that tight annotations are available only for the validation sets listed in \autoref{tbl:dataset}, which is a common setting in previous studies on learning from noisy labels~\cite{jiang2018mentornet,ren2018learning}.
The case where no ideal tight labels are available is left for future work.

\subsection{Evaluation}
Speaker diarization performance is evaluated using diarization error rate (DER), defined as the sum of the missed detection rate (MI), the false alarm rate (FA), and the speaker confusion rate (CF).
For the ASR corpora, we use labels obtained via forced alignment as a reference for DER computation, as they can be considered ideal tight labels.
For the diarization corpora, we use the provided labels as reference.

\section{Results}
\subsection{Speaker diarization}
\subsubsection{Main results}

We first present the overall results on the in-domain corpora in \autoref{tbl:results_indomain}.
For the baseline models ($\mathtt{B1}$), we use the annotated labels shown in \autoref{tbl:dataset} as supervision, whereas the topline models ($\mathtt{B2}$) use the ideally tightened labels via forced alignment for the ASR corpora, while the diarization corpora are used as is since their annotations are already tight.
We observed a substantial DER gap between the two models for the ASR corpora, while the DERs remain largely unchanged on the diarization corpora.
This trend is consistent with the prior study showing that models tend to learn the degree of label tightness specific to each corpus \cite{horiguchi2025can}.
The objective of the proposed method is to close this gap without relying on ideal tight labels.

By applying the proposed method, DER was significantly reduced for the ASR corpora, indicating that the models can successfully produce tightened outputs ($\mathtt{P1}$--$\mathtt{P3}$).
This substantially reduced the discrepancy in boundary tightness across domains and enhanced model interpretability.
This fact is also supported by the results on the out-of-domain dataset, shown in \autoref{tbl:results_dihard}.
The baseline model exhibited a high DER on the DIHARD III corpus, and the gap from the topline indicates that the model tended to produce loose speech intervals.
In contrast, the proposed method enforced consistently tight outputs, preventing predictions from being affected by unpredictable loosening behavior.
Among the three tightening methods, the SC-based approach achieved DER reductions comparable to or greater than those of the other methods.
Detailed analyses of each tightening method are provided in \autoref{sec:analysis}.

\begin{figure}[t]
\hspace{4.5em}\includegraphics{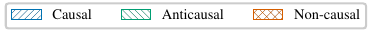}\vspace{0.3em}\\
\includegraphics[width=\linewidth]{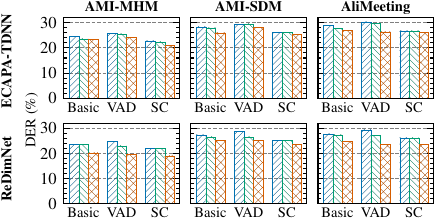}
\caption{DER (\%) of causal and anticausal models after co-training compared with non-causal models ($\mathtt{P1}$--$\mathtt{P3}$).}
\label{fig:results_causal_anticausal}
\end{figure}

We also report the performance of the causal and anticausal models obtained through co-training for each tightening method in \autoref{fig:results_causal_anticausal}.
Notably, the non-causal model trained with pseudo labels generated from the outputs of the causal and anticausal models consistently outperformed the original causal and anticausal models.
This indicates that label tightening based on causal--anticausal consistency was effective.

\begin{figure*}[t]
\begin{minipage}{0.29\linewidth}
\includegraphics[width=\linewidth]{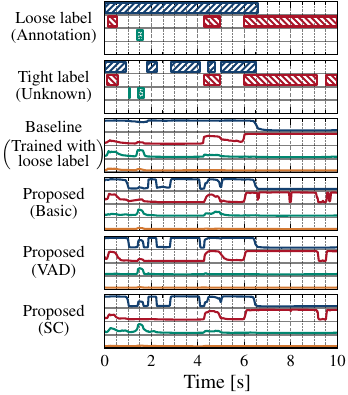}
\caption{Example of labels and processing results for a 10-second chunk.}\label{fig:sample}
\end{minipage}
\hfill
\begin{minipage}{0.32\linewidth}
\includegraphics[width=\linewidth]{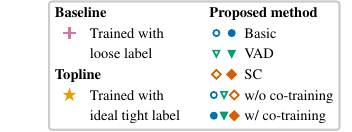}\vspace{0.3em}
\includegraphics[width=\linewidth]{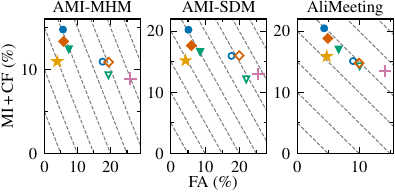}
\caption{FA vs.\ MI+CF of each method using the ReDimNet-based models. The diagonal dashed lines represent equal-DER contours, since DER=MI+FA+CF.}\label{fig:mi_facf}
\end{minipage}
\hfill
\begin{minipage}{0.35\linewidth}
\includegraphics[width=\linewidth]{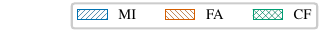}\vspace{0.3em}
\includegraphics[width=\linewidth]{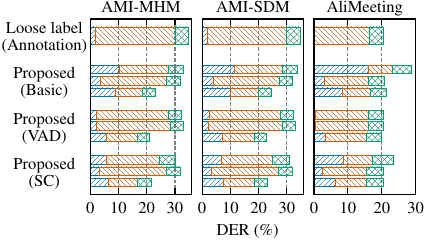}
\caption{Qualities of tightened training-set labels obtained using ReDimNet-based models measured by DER against ideal tight labels. For each proposed method, the three bars (from top to bottom) represent vanilla, with restoration, and with restoration and co-training.}\label{fig:tightened_label}
\end{minipage}
\end{figure*}

\Autoref{fig:sample} shows an example of a 10-second chunk with the corresponding loose and tight labels and the inference results using the ReDimNet-based models.
Comparing the loose label (first row) and the tight label (second row), it is clear that some speech intervals are merged in the loose label.
When a model was trained with such loose labels, its predictions also became loose (third row), whereas the proposed methods produce tighter predictions (fourth to sixth row).

\subsubsection{Analysis of tightening methods and ablation study}\label{sec:analysis}
\Autoref{fig:mi_facf} compares the tightening methods on the ASR corpora and includes an ablation study of co-training in terms of FA vs.\ MI\,+\,CF.
Without co-training, all tightening methods reduced FA at the cost of a slight increase in MI\,+\,CF, resulting in modest improvements in DERs.
With co-training, FA was reduced more substantially, yielding further improvement in DER.
However, the resulting MI\,+\,CF was higher than that of the topline models trained with tight labels.

Among the tightening methods, SC tightening achieved the best DER, whereas VAD tightening yielded the lowest MI.
This result was consistent with the discussion in \autoref{sec:vad_tightening} that VAD tightening behaves conservatively.
The impact of this over-tightening on downstream tasks is discussed in \autoref{sec:asr}.

\subsubsection{Analysis of tightened labels}
We next analyzed each corpus’s tightened training-set labels, rather than the model outputs.
The results are shown in \autoref{fig:tightened_label}.
We first examine the case without restoration in \autoref{sec:restore} or co-training in \autoref{sec:cotraining}.
Each tightening method substantially reduced FA; however, it also significantly increased MI, and in some cases, the total error even exceeded that of the original loose labels.
Introducing restoration made the reduction in FA somewhat smaller while effectively restraining the increase in MI.
Furthermore, when co-training was applied, particularly on AMI-MHM and AMI-SDM, FA was substantially reduced and the increase in MI was controlled.
This resulted in efficient tightening and a large reduction in DER.
On AliMeeting, although some FA could still be traded off for MI, DER remained almost unchanged.
Nevertheless, as shown in \autoref{fig:mi_facf}, training with tightened labels yielded lower DER than training without tightening, indicating that the proposed method still improved performance.
This may be because the MI introduced during label generation tended to occur in inherently hard-to-detect regions, thereby limiting its impact on the final DER.

\subsection{Multi-talker ASR}\label{sec:asr}
Finally, we evaluated each diarization system in two multi-talker ASR pipelines: single-channel ASR and multi-channel ASR.
Both pipelines used diarization results from AMI-SDM, obtained using the ReDimNet-based models.

\noindent\textbf{Multi-channel ASR.} 
The ASR performance was measured under the AMI multiple distant microphone (AMI-MDM) setup, which uses the recordings of an eight-channel microphone array.
Note that AMI-SDM used for diarization corresponds to the recordings from the first channel of each microphone array.
As the multi-channel ASR pipeline, we enhanced each utterance with GSS~\cite{boeddeker2018front,raj2023gpu} followed by transcription with Whisper large-v3~\cite{radford2023robust}.
Following the previous study~\cite{horiguchi2025can}, we used a morphological closing before beamforming in GSS, with the width tuned on the validation set.

\noindent\textbf{Single-channel ASR.} As a single-channel ASR pipeline, we used self-enrolled diarization-conditioned Whisper (SE-DiCoW)~\cite{polok2026sedicow}.
It directly takes a multi-talker recording with the corresponding diarization results as inputs.
We also used a morphological closing before feeding into the SE-DiCoW model.

\noindent\textbf{Results.}
In \autoref{tbl:results_asr}, we reported time-constrained minimum-permutation word error rates (tcpWERs) computed using the MeetEval toolkit with the CHiME-8 text normalizer~\cite{von2023meeteval}.
Consistent with the previous study~\cite{horiguchi2025can}, training with tight labels improved tcpWER over loose-label training.
For the proposed methods, both VAD and SC tightening achieved better results than the model trained with loose labels ($\mathtt{B1}$ vs. $\mathtt{B2}$).
In terms of DER, SC tightening ($\mathtt{P3}$) achieved the best performance as in \autoref{tbl:results_indomain}; however, an increase in MI\,+\,CF led to higher deletion errors in ASR, making VAD tightening ($\mathtt{P2}$) preferable in practice.
Incorporating downstream-task awareness into the training procedure remains an important direction for future work.

\begin{table}[t]
\renewrobustcmd{\bfseries}{\fontseries{b}\selectfont}
\renewrobustcmd{\boldmath}{}
\newrobustcmd{\B}{\bfseries}
\sisetup{detect-weight,mode=text}
\caption{tcpWER (\%) on AMI. The numbers in parentheses represent deletion, insertion, and substitution errors.}\label{tbl:results_asr}
\centering
\setlength{\tabcolsep}{3pt}
\resizebox{\linewidth}{!}{%
\begin{tabular}{@{}lccc@{\,}cc@{\,}c@{}}
\toprule
&&&\multicolumn{2}{c}{Multi-channel ASR}&\multicolumn{2}{c@{}}{Single-channel ASR}\\
&Label&Tightening&\multicolumn{2}{c}{(GSS\,+\,Whisper)}&\multicolumn{2}{c@{}}{(SE-DiCoW)}\\\midrule
\multicolumn{5}{@{}l}{\textbf{Baseline \& topline}}\\
($\mathtt{B1}$)&Loose& N/A&30.28&{\scriptsize (19.57\,/\,5.36\,/\,5.35)}&28.30&{\scriptsize (12.70\,/\,9.05\,/\,6.55)}\\
($\mathtt{B2}$)&Tight&N/A&28.57&{\scriptsize (18.85\,/\,4.66\,/\,5.17)}&25.93&{\scriptsize(11.83\,/\,7.59\,/\,6.51)}\\\midrule
\multicolumn{5}{@{}l}{\textbf{Proposed method}}\\
($\mathtt{P1}$)&Loose& Basic&31.43&{\scriptsize(20.57\,/\,5.67\,/\,5.20)}&28.21&{\scriptsize(13.51\,/\,8.47\,/\,6.23)}\\
($\mathtt{P2}$)&Loose& VAD&\B 29.44&{\scriptsize(19.00\,/\,4.80\,/\,5.63)}&\B 26.31&{\scriptsize(11.88\,/\,7.66\,/\,6.77)}\\
($\mathtt{P3}$)&Loose& SC&30.19&{\scriptsize(19.83\,/\,5.20\,/\,5.32)}&26.68&{\scriptsize(12.77\,/\,7.55\,/\,6.36)}\\
\bottomrule
\end{tabular}%
}
\end{table}

\section{Conclusions}
In this paper, we proposed a training method for speaker diarization models with loose ASR labels while enabling tight boundary inference.
By leveraging the asymmetric padding properties of causal and anticausal models, the method refines loose annotations through co-training.
This eliminates the need for costly manual annotation or access to headset microphones, making tight-boundary training applicable to more diverse datasets.

\noindent
\textbf{Limitations.}
Determining convergence in co-training still requires a small amount of ideal tight labels for validation.
Moreover, our experiments were limited to moderately sized datasets with forced-alignment-based labels.
Large-scale evaluation without tight labels remains future work.
In addition, our model was smaller than large-scale pretrained models~\cite{chen2022wavlm}, which may limit performance compared with causalized large models~\cite{zhao2022speech}.
Finally, progressive tightening sometimes degraded downstream performance, and identifying an optimal degree of tightening is another promising direction.

\clearpage
\section{Generative AI Use Disclosure}
The authors used a large language model (ChatGPT) only to assist with language editing and polishing.
All technical content was developed and verified by the authors.

\bibliographystyle{IEEEtran}
\bibliography{mybib}

@article{carletta2007unleashing,
    title={Unleashing the killer corpus: experiences in creating the multi-everything {AMI Meeting Corpus}},
    author={Carletta, Jean},
    journal={Language Resources and Evaluation},
    volume={41},
    number={2},
    pages={181--190},
    year={2007},
}

@article{chen2022wavlm,
    title={{WavLM}: Large-scale self-supervised pre-training for full stack speech processing},
    author={Chen, Sanyuan and Wang, Chengyi and Chen, Zhengyang and Wu, Yu and Liu, Shujie and Chen, Zhuo and Li, Jinyu and Kanda, Naoyuki and Yoshioka, Takuya and Xiao, Xiong and others},
    journal={IEEE Journal of Selected Topics in Signal Processing},
    volume={16},
    number={6},
    pages={1505--1518},
    year={2022},
    publisher={IEEE}
}

@article{cheng2025sequence,
    title={Sequence-to-sequence neural diarization with automatic speaker detection and representation},
    author={Cheng, Ming and Lin, Yuke and Li, Ming},
    journal={IEEE TASLPRO},
    year={2025},
    volume={33},
    pages={2719--2734}
}

@article{horiguchi2022encoderdecoder,
    title={Encoder-Decoder Based Attractors for End-to-End Neural Diarization},
    journal={IEEE/ACM TASLP},
    author={Horiguchi, Shota and Fujita, Yusuke and Watanabe, Shinji and Xue, Yawen and Garc{\'i}a, Paola},
    volume={30},
    year={2022},
    pages={1493--1507},
}

@article{nagrani2020voxceleb,
    title={{VoxCeleb}: Large-scale speaker verification in the wild},
    author={Nagrani, Arsha and Chung, Joon Son and Xie, Weidi and Zisserman, Andrew},
    journal={Computer Speech \& Language},
    volume={60},
    pages={101027},
    year={2020}
}

@inproceedings{barker2018fifth,
    title={The fifth {`CHiME' Speech Separation and Recognition Challenge}: dataset, task and baselines},
    author={Barker, Jon and Watanabe, Shinji and Vincent, Emmanuel and Trmal, Jan},
    booktitle={Proc. Interspeech},
    pages={1561--1565},
    year={2018}
}

@inproceedings{boeddeker2018front,
    author={Boeddeker, Christoph and Heitkaemper, Jens and Schmalenstoeer, Joerg and Drude, Lukas and Heymann, Jahn and Haeb-Umbach, Reinhold},
    title={Front-End Processing for the {CHiME-5} Dinner Party Scenario},
    booktitle={Proc. CHiME-5},
    pages={35--40},
    year={2018}
}

@inproceedings{bredin2023pyannote,
    title={{pyannote.audio 2.1} speaker diarization pipeline: principle, benchmark, and recipe},
    author={Bredin, Herv{\'e}},
    booktitle={Proc. Interspeech},
    pages={1983--1987},
    year={2023}
}

@inproceedings{chung2020spot,
    title={Spot the conversation: Speaker diarisation in the wild},
    author={Chung, Joon Son and Huh, Jaesung and Nagrani, Arsha and Afouras, Triantafyllos and Zisserman, Andrew},
    year={2020},
    booktitle={Proc. Interspeech},
    pages={299--303}
}

@inproceedings{desplanques2020ecapatdnn,
    author={Brecht Desplanques and Jenthe Thienpondt and Kris Demuynck},
    title={{ECAPA-TDNN}: Emphasized Channel Attention, Propagation and Aggregation in {TDNN} Based Speaker Verification},
    year={2020},
    booktitle={Proc. Interspeech},
    pages={3830--3834},
}

@inproceedings{ebbers2020forward,
    title={Forward-Backward Convolutional Recurrent Neural Networks and Tag-Conditioned Convolutional Neural Networks for Weakly Labeled Semi-Supervised Sound Event Detection},
    author={Ebbers, Janek and Haeb-Umbach, Reinhold},
    booktitle={Proc. DCASE},
    pages={41--45},
    year={2020}
}

@inproceedings{fu2021aishell,
    title={{AISHELL-4}: An open source dataset for speech enhancement, separation, recognition and speaker diarization in conference scenario},
    author={Fu, Yihui and Cheng, Luyao and Lv, Shubo and Jv, Yukai and Kong, Yuxiang and Chen, Zhuo and Hu, Yanxin and Xie, Lei and Wu, Jian and Bu, Hui and Xu, Xin and Du, Jun and Chen, Jingdong},
    booktitle={Proc. Interspeech},
    year={2021},
    pages={3665--3669}
}

@inproceedings{fujita2019end1,
    author={Fujita, Yusuke and Kanda, Naoyuki and Horiguchi, Shota and Nagamatsu, Kenji and Watanabe, Shinji},
    title={End-to-End Neural Speaker Diarization with Permutation-Free Objectives},
    booktitle={Proc. Interspeech},
    year={2019},
    pages={4300-4304}
}

@inproceedings{fujita2019end2,
    author={Fujita, Yusuke and Kanda, Naoyuki and Horiguchi, Shota and Xue, Yawen and Nagamatsu, Kenji and Watanabe, Shinji},
    title={End-to-End Neural Speaker Diarization with Self-attention},
    booktitle={Proc. ASRU},
    year={2019},
    pages={296-303}
}

@inproceedings{ghosh2025audio,
    title={Audio {Flamingo} 3: Advancing Audio Intelligence with Fully Open Large Audio Language Models},
    author={Ghosh, Sreyan and Goel, Arushi and Kim, Jaehyeon and Kumar, Sonal and Kong, Zhifeng and Lee, Sang-gil and Yang, Chao-Han Huck and Duraiswami, Ramani and Manocha, Dinesh and Valle, Rafael and others},
    booktitle={Proc. NeurIPS},
    year={2025}
}

@inproceedings{guo2018curriculumnet,
    title={{CurriculumNet}: Weakly supervised learning from large-scale web images},
    author={Guo, Sheng and Huang, Weilin and Zhang, Haozhi and Zhuang, Chenfan and Dong, Dengke and Scott, Matthew R and Huang, Dinglong},
    booktitle={Proc. ECCV},
    pages={135--150},
    year={2018}
}

@inproceedings{han2018co,
    title={Co-teaching: Robust training of deep neural networks with extremely noisy labels},
    author={Han, Bo and Yao, Quanming and Yu, Xingrui and Niu, Gang and Xu, Miao and Hu, Weihua and Tsang, Ivor and Sugiyama, Masashi},
    booktitle={Proc. NeurIPS},
    volume={31},
    pages={8527--8537},
    year={2018}
}

@inproceedings{han2025fine,
    title={Fine-tune Before Structured Pruning: Towards Compact and Accurate Self-Supervised Models for Speaker Diarization},
    author={Han, Jiangyu and Landini, Federico and Rohdin, Johan and Silnova, Anna and Diez, Mireia and Cernocky, Jan and Burget, Lukas},
    booktitle={Proc. Interspeech},
    year={2025},
    pages={1583--1587}
}

@inproceedings{han2025leveraging,
    title={Leveraging Self-Supervised Learning for Speaker Diarization},
    author={Han, Jiangyu and Landini, Federico and Rohdin, Johan and Silnova, Anna and Diez, Mireia and Burget, Lukas},
    booktitle={Proc. ICASSP},
    year={2025}
}

@inproceedings{harkonen2024eend,
    title={{EEND-M2F}: Masked-attention mask transformers for speaker diarization},
    author={H{\"a}rk{\"o}nen, Marc and Broughton, Samuel J and Samarakoon, Lahiru},
    booktitle={Proc. Interspeech},
    pages={37--41},
    year={2024}
}

@inproceedings{horiguchi2025can,
    title={Can We Really Repurpose Multi-Speaker {ASR} Corpus for Speaker Diarization?},
    author={Horiguchi, Shota and Tawara, Naohiro and Ashihara, Takanori and Ando, Atsushi and Delcroix, Marc},
    booktitle={Proc. ASRU},
    year={2025}
}

@inproceedings{horiguchi2025pretraining,
    title={Pretraining Multi-Speaker Identification for Neural Speaker Diarization},
    author={Horiguchi, Shota and Ando, Atsushi and Delcroix, Marc and Tawara, Naohiro},
    booktitle={Proc. Interspeech},
    pages={1608--1612},
    year={2025}
}

@inproceedings{hu2018squeeze,
    title={Squeeze-and-excitation networks},
    author={Hu, Jie and Shen, Li and Sun, Gang},
    booktitle={Proc. CVPR},
    pages={7132--7141},
    year={2018}
}

@inproceedings{jiang2018mentornet,
    title={{MentorNet}: Learning data-driven curriculum for very deep neural networks on corrupted labels},
    author={Jiang, Lu and Zhou, Zhengyuan and Leung, Thomas and Li, Li-Jia and Fei-Fei, Li},
    booktitle={Proc. ICML},
    pages={2304--2313},
    year={2018}
}

@inproceedings{kalda2022collar,
    title={Collar-Aware Training for Streaming Speaker Change Detection in Broadcast Speech},
    author={Kalda, Joonas and Alum{\"a}e, Tanel},
    booktitle={Proc. Odyssey 2022},
    pages={141--147},
    year={2022}
}

@inproceedings{kingma2015adam,
    author={Kingma, Diederik P. and Ba, Jimmy},
    booktitle={Proc. ICLR},
    title={Adam: A Method for Stochastic Optimization},
    year={2015}
}

@inproceedings{kinoshita2021integrating,
    author={Kinoshita, Keisuke and Delcroix, Marc and Tawara, Naohiro},
    title={Integrating End-to-End Neural and Clustering-Based Diarization: Getting the Best of Both Worlds},
    booktitle={Proc. ICASSP},
    pages={7198--7202},
    year={2021}
}

@inproceedings{liu2022msdwild,
    title={{MSDWild}: Multi-modal Speaker Diarization Dataset in the Wild},
    author={Liu, Tao and Fan, Shuai and Xiang, Xu and Song, Hongbo and Lin, Shaoxiong and Sun, Jiaqi and Han, Tianyuan and Chen, Siyuan and Yao, Binwei and Liu, Sen and Wu, Yifei and Qian, Yanmin and Yu, Kai},
    booktitle={Proc. Interspeech},
    pages={1476--1480},
    year={2022}
}

@inproceedings{lukasik2020does,
    title={Does label smoothing mitigate label noise?},
    author={Lukasik, Michal and Bhojanapalli, Srinadh and Menon, Aditya and Kumar, Sanjiv},
    booktitle={Proc. ICML},
    pages={6448--6458},
    year={2020},
    organization={PMLR}
}

@inproceedings{medennikov2020targetspeaker,
    title={Target-Speaker Voice Activity Detection: a Novel Approach for Multi-Speaker Diarization in a Dinner Party Scenario},
    author={Medennikov, Ivan and Korenevsky, Maxim and Prisyach, Tatiana and Khokhlov, Yuri and Korenevskaya, Mariya and Sorokin, Ivan and Timofeeva, Tatiana and Mitrofanov, Anton and Andrusenko, Andrei and Podluzhny, Ivan and Laptev, Aleksandr and Romanenko, Aleksei},
    year={2020},
    pages={274--278},
    booktitle={Proc. Interspeech},
}

@inproceedings{nagrani2017voxceleb,
    title={{VoxCeleb}: A large-scale speaker identification dataset},
    author={Nagrani, Arsha and Chung, Joon Son and Zisserman, Andrew},
    booktitle={Proc. Interspeech},
    pages={2616--2620},
    year={2017}
}

@inproceedings{natarajan2013learning,
    title={Learning with noisy labels},
    author={Natarajan, Nagarajan and Dhillon, Inderjit S and Ravikumar, Pradeep K and Tewari, Ambuj},
    booktitle={Proc. NeurIPS},
    volume={26},
    pages={1196--1204},
    year={2013}
}

@inproceedings{ohashi2025towards,
    title={Towards a Japanese Full-duplex Spoken Dialogue System},
    author={Ohashi, Atsumoto and Iizuka, Shinya and Jiang, Jingjing and Higashinaka, Ryuichiro},
    year={2025},
    booktitle={Proc. Interspeech},
    pages={1783--1787},
}

@inproceedings{patrini2017making,
    title={Making deep neural networks robust to label noise: A loss correction approach},
    author={Patrini, Giorgio and Rozza, Alessandro and Krishna Menon, Aditya and Nock, Richard and Qu, Lizhen},
    booktitle={Proc. CVPR},
    pages={1944--1952},
    year={2017}
}

@inproceedings{plaquet2023powerset,
    title={Powerset multi-class cross entropy loss for neural speaker diarization},
    author={Plaquet, Alexis and Bredin, Herv{\'e}},
    booktitle={Proc. Interspeech},
    pages={3222--3226},
    year={2023}
}

@inproceedings{plaquet2025mamba,
    title={Mamba-based Segmentation Model for Speaker Diarization},
    author={Plaquet, Alexis and Tawara, Naohiro and Delcroix, Marc and Horiguchi, Shota and Ando, Atsushi and Araki, Shoko},
    booktitle={Proc. ICASSP},
    year={2025}
}

@inproceedings{polok2026sedicow,
    title={{SE-DiCoW}: Self-Enrolled Diarization-Conditioned {Whisper}},
    author={Polok, Alexander and Klement, Dominik and Cornell, Samuele and Wiesner, Matthew and {\v{C}}ernock{\`y}, Jan and Khudanpur, Sanjeev and Burget, Luk{\'a}{\v{s}}},
    booktitle={Proc. ICASSP},
    year={2026}
}

@inproceedings{radford2023robust,
    title={Robust speech recognition via large-scale weak supervision},
    author={Radford, Alec and Kim, Jong Wook and Xu, Tao and Brockman, Greg and McLeavey, Christine and Sutskever, Ilya},
    booktitle={Proc. ICML},
    pages={28492--28518},
    year={2023}
}

@inproceedings{raj2023gpu,
    title={{GPU}-accelerated guided source separation for meeting transcription},
    author={Raj, Desh and Povey, Daniel and Khudanpur, Sanjeev},
    booktitle={Proc. Interspeech},
    pages={3507--3511},
    year={2023}
}

@inproceedings{ren2018learning,
    title={Learning to reweight examples for robust deep learning},
    author={Ren, Mengye and Zeng, Wenyuan and Yang, Bin and Urtasun, Raquel},
    booktitle={Proc. ICML},
    pages={4334--4343},
    year={2018},
}

@inproceedings{ryant2019second,
    title={{The Second DIHARD Diarization Challenge}: Dataset, Task, and Baselines},
    author={Ryant, Neville and Church, Kenneth and Cieri, Christopher and Cristia, Alejandrina and Du, Jun and Ganapathy, Sriram and Liberman, Mark},
    booktitle={Proc. Interspeech},
    pages={978--982},
    year={2019}
}

@inproceedings{ryant2021third,
    author={Ryant, Neville and Singh, Prachi and Krishnamohan, Venkat and Varma, Rajat and Church, Kenneth and Cieri, Christopher and Du, Jun and Ganapathy, Sriram and Liberman, Mark},
    title={The Third {DIHARD} Diarization Challenge},
    year={2021},
    booktitle={Proc. Interspeech},
    pages={3570--3574},
}

@inproceedings{van2020dipco,
    title={{DiPCo}---Dinner Party Corpus},
    author={Van Segbroeck, Maarten and Zaid, Ahmed and Kutsenko, Ksenia and Huerta, Cirenia and Nguyen, Tinh and Luo, Xuewen and Hoffmeister, Bj{\"o}rn and Trmal, Jan and Omologo, Maurizio and Maas, Roland},
    year={2020},
    booktitle={Proc. Interspeech},
    pages={434--436}
}

@inproceedings{veluri2024beyond,
    title={Beyond Turn-Based Interfaces: Synchronous {LLM}s as Full-Duplex Dialogue Agents},
    author={Veluri, Bandhav and Peloquin, Benjamin and Yu, Bokai and Gong, Hongyu and Gollakota, Shyamnath},
    booktitle={Proc. NAACL},
    pages={21390--21402},
    year={2024}
}

@inproceedings{von2023meeteval,
    title={{MeetEval}: A toolkit for computation of word error rates for meeting transcription systems},
    author={von Neumann, Thilo and Boeddeker, Christoph and Delcroix, Marc and Haeb-Umbach, Reinhold},
    booktitle={Proc. CHiME},
    pages={27--32},
    year={2023}
}

@inproceedings{wang2023target,
    title={Target speaker voice activity detection with transformers and its integration with end-to-end neural diarization},
    author={Wang, Dongmei and Xiao, Xiong and Kanda, Naoyuki and Yoshioka, Takuya and Wu, Jian},
    booktitle={Proc. ICASSP},
    year={2023}
}

@inproceedings{yakovlev2024reshape,
    title={Reshape dimensions network for speaker recognition},
    author={Yakovlev, Ivan and Makarov, Rostislav and Balykin, Andrei and Malov, Pavel and Okhotnikov, Anton and Torgashov, Nikita},
    booktitle={Proc. Interspeech},
    pages={3235--3239},
    year={2024},
}

@inproceedings{yu2019does,
    title={How does Disagreement Help Generalization against Label Corruption?},
    author={Yu, Xingrui and Han, Bo and Yao, Jiangchao and Niu, Gang and Tsang, Ivor and Sugiyama, Masashi},
    booktitle={Proc. ICML},
    pages={7164--7173},
    year={2019}
}

@inproceedings{yu2021dual,
    title={Dual-mode {ASR}: Unify and improve streaming {ASR} with full-context modeling},
    author={Yu, Jiahui and Han, Wei and Gulati, Anmol and Chiu, Chung-Cheng and Li, Bo and Sainath, Tara N and Wu, Yonghui and Pang, Ruoming},
    booktitle={Proc. ICLR},
    year={2021}
}

@inproceedings{yu2022m2met,
    author={Yu, Fan and Zhang, Shiliang and Fu, Yihui and Xie, Lei and Zheng, Siqi and Du, Zhihao and Huang, Weilong and Guo, Pengcheng and Yan, Zhijie and Ma, Bin and Xu, Xin and Bu, Hui},
    booktitle={Proc. ICASSP},
    title={{M2MeT}: The {ICASSP} 2022 Multi-Channel Multi-Party Meeting Transcription Challenge},
    year={2022},
    pages={6167-6171},
}

@inproceedings{zeghidour2021dive,
    title={{DIVE}: End-to-End Speech Diarization via Iterative Speaker Embeddings},
    author={Zeghidour, Neil and Teboul, Olivier and Grangier, David},
    booktitle={Proc. ASRU},
    pages={702--709},
    year={2021}
}

@inproceedings{zhao2022speech,
    title={Speech enhancement using self-supervised pre-trained model and vector quantization},
    author={Zhao, Xiao-Ying and Zhu, Qiu-Shi and Zhang, Jie},
    booktitle={Proc. APSIPA ASC},
    pages={330--334},
    year={2022},
}

@misc{cui2026turnguide,
    title={{TurnGuide}: Enhancing meaningfull full duplex spoken interactions via dynamic turn-level text-speech interleaving},
    author={Cui, Wenqian and Zhu, Lei and Li, Xiaohun and Gui, Zhihan},
    howpublished={arxiv:2508.07375},
    year={2026}
}

@misc{defossez2024moshi,
    title={Moshi: a speech-text foundation model for real-time dialogue},
    author={D{\'e}fossez, Alexandre and Mazar{\'e}, Laurent and Orsini, Manu and Royer, Am{\'e}lie and P{\'e}rez, Patrick and J{\'e}gou, Herv{\'e} and Grave, Edouard and Zeghidour, Neil},
    howpublished={arXiv:2410.00037},
    year={2024}
}

@misc{fu2024investigating,
    title={Investigating the Effects of Large-Scale Pseudo-Stereo Data and Different Speech Foundation Model on Dialogue Generative Spoken Language Model},
    author={Fu, Yu-Kuan and Lee, Cheng-Kuang and Wang, Hsiu-Hsuan and Lee, Hung-yi},
    howpublished={arXiv:2407.01911},
    year={2024}
}

@misc{han2025efficient,
    title={Efficient and generalizable speaker diarization via structured pruning of self-supervised models},
    author={Han, Jiangyu and P{\'a}lka, Petr and Delcroix, Marc and Landini, Federico and Rohdin, Johan and Cernock{\`y}, Jan and Burget, Luk{\'a}{\v{s}}},
    howpublished={arXiv:2506.18623},
    year={2025}
}

@misc{ryant2018first,
    author={Ryant, Neville and Church, Kenneth and Cieri, Christopher and Cristia, Alejandrina and Du, Jun and Ganapathy, Sriram and Liberman, Mark},
    title = {First {DIHARD} challenge evaluation plan},
    year = {2018},
    howpublished={https://zenodo.org/record/1199638},
}

@misc{ryant2020third,
    title={Third {DIHARD} challenge evaluation plan},
    author={Ryant, Neville and Church, Kenneth and Cieri, Christopher and Du, Jun and Ganapathy, Sriram and Liberman, Mark},
    howpublished={arXiv:2006.05815},
    year={2020}
}

\end{document}